\documentclass[runningheads]{svmult}

\usepackage{makeidx}   
\usepackage{graphicx}  
\usepackage{subeqnar}  
\usepackage{multicol}  
\usepackage{physprbb}  
\makeindex             

\begin{document}
\title*{Studying the Dynamics of Star Forming and IR Luminous Galaxies with
Infrared Spectroscopy}
\toctitle{Studying the Dynamics of Star Forming and IR Luminous Galaxies with
Infrared Spectroscopy}
%
%
\titlerunning{Galaxy Dynamics with IR Spectroscopy}
%
\author{Reinhard Genzel
\and Linda J.~Tacconi
\and Marco Barden
\and Matthew D.~Lehnert
\and Dieter Lutz
\and Dimitra Rigopoulou
\and Niranjan Thatte}
\tocauthor{R.~Genzel, L.J.~Tacconi, M.~Barden, M.D.~Lehnert, D.~Lutz, 
D.~Rigopoulou, N.~Thatte}
\authorrunning{R. Genzel et al.}
%
%
\institute{Max-Planck Institut f\"{u}r extraterrestrische Physik, 
Garching, FRG}

\maketitle              

\begin{abstract}
With the advent of efficient near-IR spectrometers on 10m-class telescopes,
exploiting the new generation of low readout noise, large format detectors,
OH avoidance and sub-arcsecond seeing, 1-2.4$\mu m$ spectroscopy can now be
exploited for detailed galaxy dynamics and for studies of high-z galaxies.
In the following we present the results of three recent IR spectroscopy
studies on the dynamics of ULIRG mergers, on super star clusters in the
Antennae, and on the properties of the rotation curves of z$\sim$1
disk galaxies, carried out with ISAAC on the VLT, and NIRSPEC on the Keck.
\end{abstract}

\section{Ultra-Luminous Infrared Galaxies:\protect\newline 
Ellipticals and QSOs in Formation?}
\begin{figure}
\begin{center}
\includegraphics[width=10cm, bb = 48 68 542 646, clip=true]{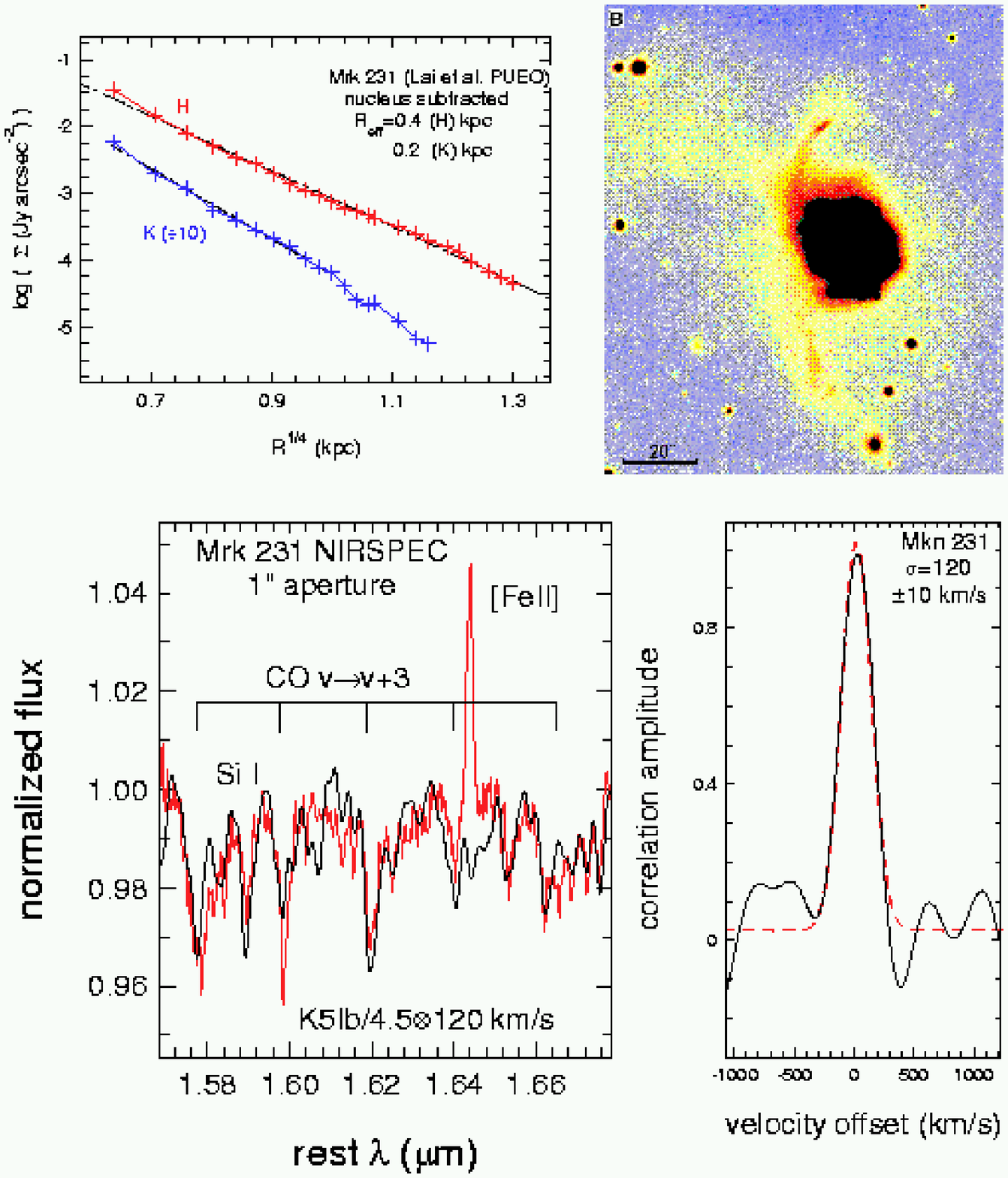}
\end{center}
\caption[]{Observations of the ULIRG/QSO Mkn231 (z=0.04), 
L$_{IR}$ = $3 \times 10^{12}$ L$_\odot$). 
Upper left: H- (red) and K-band (blue) 
surface brightness
distributions ($\sim 0.14''$ resolution) after removal of the nucleus,
from the adaptive optics observations of [19]. Upper right:
B-band image of the outer parts of Mkn~231, showing the two tidal tails, and
the central compact body of the galaxy [5]. Lower
left: H-band spectrum of the central region (red, excluding central 
$\sim 0.8''$), obtained with NIRSPEC on Keck [31]. The black
curve shows a late type star template spectrum, convolved with the best
fitting Gaussian (right inset) and diluted by a factor of 4.5. The strongest
CO and SiI stellar absorption features are marked, as is the [FeII] emission
line. Right inset: Line of sight velocity distribution of the stars
(continuous black), as derived from a Fourier quotient analysis of the
spectrum in the left inset. The best fitting Gaussian (red dashed) has a
dispersion of 120 km/s}
\label{eps1}
\end{figure}

Recent deep mid-IR [8] and submillimeter [4]
surveys have discovered a population of distant, dust enshrouded starbursts
that may contribute about half of the cosmic star formation activity at 
z$\ge$1 [13]. These IR-luminous starbursts may be
large bulges/ellipticals in formation. (Ultra)-luminous infrared galaxies
((U)LIRGs: L(1-1000$\mu $m)$\ge 10^{11}(\ge $10$^{12}$)L$_\odot$,
[24]) may be the local-Universe analogues of the high-z
population. Almost all ULIRGs are advanced mergers of gas rich disk
galaxies. Following the `ellipticals through mergers scenario' of Toomre and
Toomre [33], Kormendy and Sanders [17] proposed that ULIRGs may evolve
into ellipticals through merger induced, dissipative collapse. In the
process, such mergers may go through a very luminous starburst phase and
later evolve into classical QSOs [25]. Studies of local
ULIRGs thus may be a key to better understand the properties and evolution
of the high-z population.

To test the `ellipticals in formation' and `QSOs in formation' scenarios we
have recently begun a program of determining the fundamental structural and
dynamical properties of the stellar hosts of ULIRGs [14, 31].
If ULIRGs evolve into ellipticals, late stage ULIRG
merger remnants should lie on or near the fundamental plane ($\log \sigma
-\log r_{eff}-\mu _{eff}$) of early type galaxies. If they also evolve into
QSOs, the hosts and central black holes of ULIRGs and QSOs of similar
luminosities should have similar properties. ULIRG mergers are not in
equilibrium. Recent numerical simulations have shown, however, that because
of the rapid action of violent relaxation and tidal torques, the dynamical
properties of late stage, compact merger remnants are already fairly close
to their final equilibrium values [21, 2]. Our
project required infrared imaging and spectroscopy since ULIRGs are highly
obscured (A$_{V}$(screen)$\sim$10-40). It required 10m class
telescopes because high quality (SNR$\sim$100) spectra are necessary
to reliably extract velocity dispersions.

We selected our program ULIRGs from the BGS, 2 Jy and 1 Jy IRAS catalogs,
culling from these catalogs those sources that have single nuclei, or are
compact ($\leq $a few kpc) double nuclei systems on near-IR images, but have
definite signatures for a recent merger, such as tidal tails. We picked
sources with redshifts $\leq $0.16, where reasonably strong stellar
absorption features fall in the J, H, or K-bands. We have presently data on
18 ULIRG merger remnants, 8 of which contain QSO-like active galactic
nuclei. As an example, Figure 1 shows our Keck NIRSPEC data of Mkn~231, along
with a B-band image from [5], and the nucleus
subtracted, H/K surface brightness distribution from [19]. Mkn~231 is 
the most luminous ULIRG within z=0.05 (logL$_{IR}=12.5$) and is also
an IR-excess BAL QSO. The two tidal tails and its compact structure with a
single bright nucleus suggest it is a late stage merger of near equal mass
disk galaxies. The highly AGN diluted central spectrum can be reasonably
well fit by a Gaussian velocity dispersion of 120 km/s, which is the lowest
of our entire ULIRG sample.

\begin{figure}[h]
\begin{center}
\includegraphics[width=9cm, bb = 33 245 555 644, clip=true]{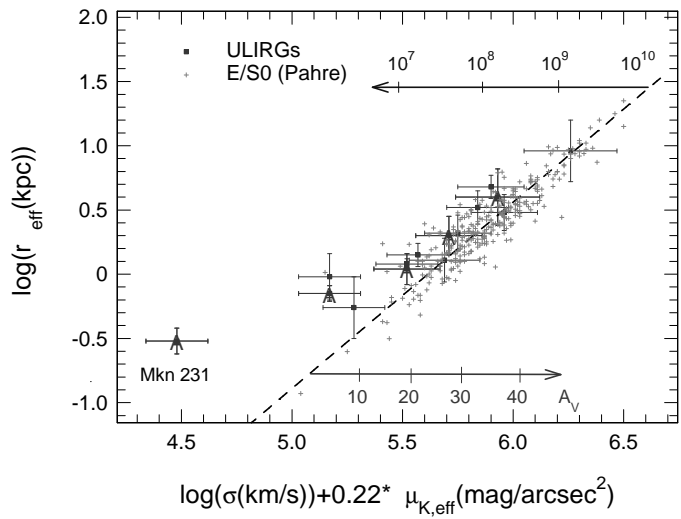}
\end{center}
\caption[]{Distribution of ULIRGs (dark symbols with 1$\sigma $ error bars,
sources dominated by AGNs marked with symbol `A') and nearby ellipticals
([22], grey crosses) in the fundamental plane, as viewed from a
projection perpendicular to the plane (best fit given by dashed line). The
direction and magnitudes of various amounts of reddening and different
stellar ages (relative to a $>$10 Gyr population) are shown at
the top and bottom of the diagram. Obscuration and younger populations tend
to cancel each other}
\label{eps2}
\end{figure}

Figure 2 shows an edge-on projection of the fundamental plane of early type
galaxies. Small grey crosses denote ellipticals/S0s from the compilation 
of [22].
Large symbols (with error bars) mark the locations of our ULIRGs,
with the AGN dominated sources marked with the letter `A'. ULIRGs fall
remarkably close to the fundamental plane. Their stellar kinematics is
largely pressure supported and in most cases stellar rotation, if present,
is smaller than the rotation of the gaseous component of the same system.
They populate a wide range of the plane, are on average similar to moderate
mass, disky ellipticals, but are well offset from giant ellipticals. Their
typical effective radii (a few kpc) and velocity dispersions ($\langle 
\sigma _{ULIRG} \rangle = 186\pm 15$ km/s) are comparable to the 
parameters of L* disky
ellipticals. In contrast giant ellipticals have effective radii r$_{eff} >
10$ kpc and an average velocity dispersion of $\langle \sigma _{gEs} \rangle 
=269\pm 7$ km/s. Late stage ULIRG mergers also show significant rotation 
($\langle v_{rot}/\sigma \rangle \ge 0.4$), and
are found in low density environments,
again similar to disky ellipticals and unlike giant ellipticals. All 117
ULIRGs of the 1 Jy catalog [16] are found in the field (or
possibly small groups), none in clusters. For comparison, of the giant
ellipticals in the [1] and [9] samples,
50\% reside in clusters of Abell richness class $\ge 0$. Our
observations thus strongly support the `ellipticals in formation' scenario
but indicate that ULIRG remnants will evolve into moderate mass disky
ellipticals, or lenticulars. Giant ellipticals must have a different
formation path (see also [9]).

The close proximity of the ULIRG mergers to the fundamental plane is
surprising. ULIRGs are heavily obscured and contain a population of young
stars. While the $\sigma $-coordinate is relatively insensitive to 
extinction effects,
the K-band surface brightness and even the effective radius are affected by
absolute and differential extinction and by population effects. The arrows
in Figure 2 show how a point in the fundamental plane would move with
extinction and aging of the underlying population. In addition the NICMOS
images of [27] show that extinction increases toward the
nuclei in most ULIRGs, resulting in effective radii decreasing with
wavelength (see Figure 1). It is thus important to determine effective radii
from near-IR images. Typical ULIRGs have A$_{V}(screen)\sim 10-50$
mag and a stellar population with a characteristic age of a few hundred Myr
[31], the combination of which approximately cancel each
other in Figure 2. So the good agreement of the ULIRGs with the fundamental
plane in the surface brightness coordinate is to some extent the result of a
`cosmic conspiracy'.

\begin{figure}[t]
\begin{center}
\includegraphics[width=10cm, bb = 60 252 550 578, clip=true]{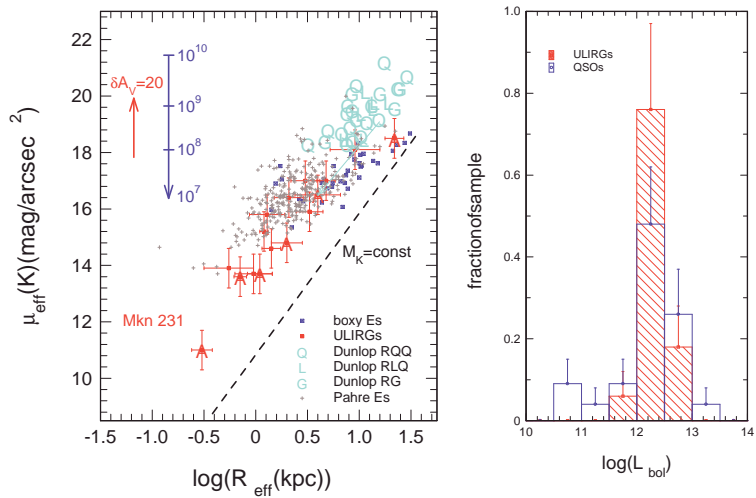}
\end{center}
\caption[]{Left inset: ULIRGs (red symbols, sources dominated by AGNs marked
with letter `A') and ellipticals (grey crosses, [22]) in the $\mu
_{eff}(K)-r_{eff}$ projection of the fundamental plane. Boxy giant
ellipticals [1, 9] are marked as
dark blue filled squares. Radio quiet (letter `Q'), radio loud (letter `L')
QSOs and radio galaxies (letter `G') [6] are in light
blue. Right inset: distributions of nuclear bolometric luminosity for the
Dunlop et al.~[6] sample (blue, unfilled histogram) and of the ULIRG
luminosities (red hatched histogram). In the case of the Dunlop et al.~
sources we multiplied the nuclear R-band luminosity be a factor 14, for the
ULIRGs we used the infrared luminosity, corrected upwards by 25\% to take
into account the contribution of the optical/UV bands}
\label{eps3}
\end{figure}

In the scenario of [25], the dusty ULIRGs are initially
powered by star formation, but as the merger progresses, accretion onto the
central black hole(s) increasingly dominates the bolometric luminosity. As
the dust and gas is cleared out in the last stages of the merger, the dusty
(infrared excess) QSO evolves into a classical, optically bright QSO. If
this scenario is correct, late stage ULIRGs (likely AGN dominated) and
optically selected QSOs of similar luminosities should have similar host
properties. To test this hypothesis, we have compared our ULIRGs to the
sample of 33 radio quiet and radio loud QSOs, and radio galaxies studied in
detail by Dunlop and coworkers (e.g. [6] and references
therein). The Dunlop AGNs sample a similar redshift range and luminosity
distribution (Figure 3) as the ULIRGs. Figure 3 shows that these 
QSOs/radio galaxies occupy the upper right right of the $\mu _{eff}$ - 
r$_{eff}$ projection of the plane, coincident with the locus of giant
ellipticals. The Dunlop et al.~AGNs, which are fairly representative of 
optically selected, local Universe QSOs/radio galaxies, have effective radii
$>$10 kpc, and 80\% of them reside in clusters. While there are no
measurements of velocity dispersions in the Dunlop AGNs yet, [3]
have compiled velocity dispersions for 73 nearby radio galaxies. The
sample average velocity dispersion of those radio galaxies is 256 km/s, in
good agreement with the hypothesis that most local Universe QSOs/radio
galaxies are large, massive early type galaxies. Some of the 
size difference between ULIRGs and QSOs may be attributable to the
nuclear concentration of bright young star forming regions in the ULIRGs
that will fade as
the population ages, and thus lead to an increase of effective radius 
with time. However, the difference in velocity dispersions
between the ULIRGs and the radio galaxies and giant ellipticals
should be unaffected by population effects. We thus conclude that the
average ULIRG host cannot evolve into a host of a typical QSO
or radio galaxy.

After correction for a mean sample K-band extinction of A$_{K}$(screen)=0.7
the average ULIRG has an absolute K-magnitude of -25.8 (5 L*), similar to
that of the radio quiet QSOs. However, the average total dynamical mass of
the ULIRG sample is 1.2$\times$10$^{11}$ M$_\odot$, or about 0.9 m*. 
Hence the L/m
ratio of the average ULIRG host is about 6 times greater than that of an old
early type galaxy, corresponding to an effective age of the stellar
population of a few hundred Myr. Taking the local black hole mass to bulge
mass ratio [10, 12],
we find that the average black hole mass in our ULIRGs is 
7$\times$10$^7$ M$_\odot$. 
In contrast Dunlop et al.\ estimate an average black hole
mass of their sample to be 1.3$\times$10$^9$ M$_\odot$. Since the nuclear
luminosities of ULIRGs and QSOs are similar (Figure 3), ULIRGs must accrete
at $>$50\% the Eddington rate, rather than the $\sim$10\%
efficiency estimated for the Dunlop QSO sample.

In summary, we conclude that ULIRGs have less massive hosts than optically
selected, low-z QSOs or radio galaxies. ULIRGs live in lower density
environments. Their black holes are more akin to Seyfert galaxies.
Nevertheless they can attain QSO-like bolometric and near-IR luminosities
because they accrete more (and more efficiently) and have younger and thus
brighter hosts. Once the merger induced, enormous influx of matter onto the
central black hole(s) ceases, ULIRGs will become inactive, moderately
massive field ellipticals or, if their black holes are fed, into objects akin
to the hard-X ray luminous, early type galaxies found recently by Chandra.
The average ULIRG cannot evolve into a classical optically selected QSO.

\section{What Is the Nature of the Super Star Clusters\protect\newline 
in the Antennae Merger?}

During the last years, many interacting and merging galaxies were discovered
to contain large numbers of very luminous young star clusters (e.g.
[15, 36]. Their overall spectral
properties suggest that they may be the progenitors of the globular cluster
populations seen in normal nearby ellipticals and spirals (e.g.~[37, 11, 26]).

\begin{figure}[p]
\begin{center}
\includegraphics[width=10cm]{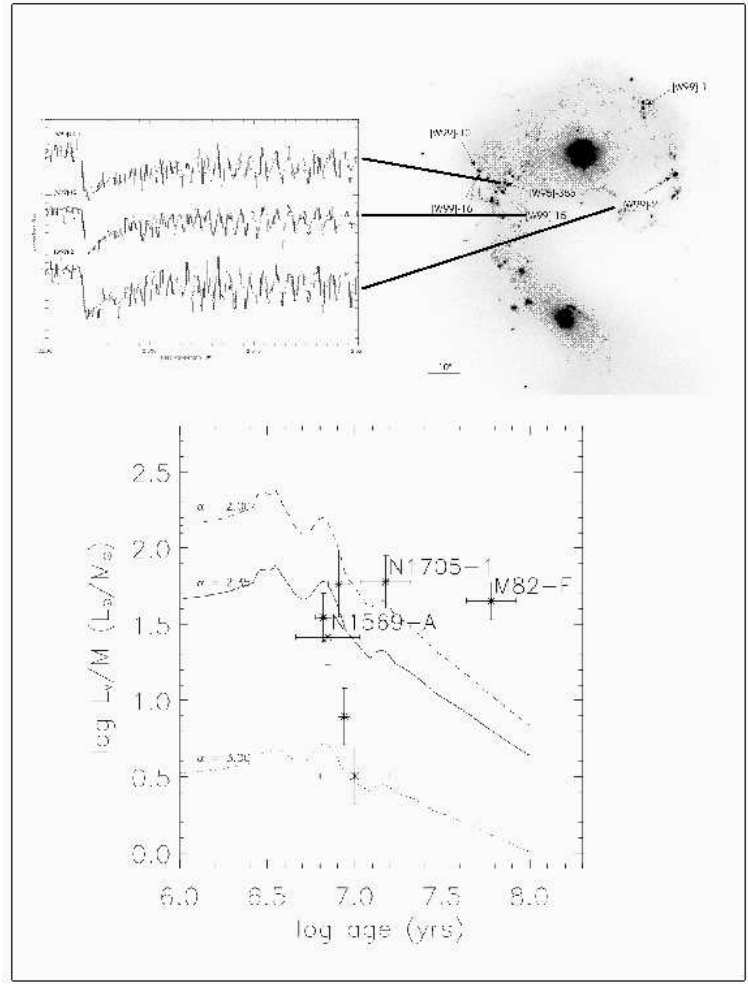}
\end{center}
\caption[]{Observations of the dynamics of super star clusters in NGC~
4038/39. Upper right: Ks-band image of the merger system, with clusters
observed with ISAAC and UVES marked. Upper left: R=9500 ISAAC spectra of
three of the clusters, along with the best fitting stellar template (red).
Bottom: V-band luminosity to mass ratio as a function of age, for 4 of the
Antennae clusters [20], along with a cluster in M82, and two
clusters in the blue dwarfs NGC 1705 and NGC 1569 [29, 30].
The different curves are IMFs with different power law
slopes, over a mass range from 0.1 to 100 M$_\odot$}
\label{eps4}
\end{figure}

To test the `globular clusters in formation' scenario and determine cluster
masses and lifetimes, we observed several of the brightest young clusters in
the Antennae (NGC4038/39) merger system with ISAAC and UVES high resolution
spectroscopy [20]. Figure 4 marks the observed clusters on a
K-band image. The target clusters were selected to be bright and exhibit
large equivalent widths of 2.3$\mu $m 0-2 CO stellar absorption, indicative
of the presence of late type supergiants. The left upper inset of Figure 4
shows ISAAC spectra of the 0-2 CO overtone bands for the three clusters
observed with ISAAC ($\sigma _{instr}(FWHM)=14 km/s)$, along with the best
fit, Gaussian broadened stellar templates. In the case of UVES ($\sigma
_{instr}(FWHM)=3.4 km/s)$ we targeted the CaT and other metal absorption
lines in the 8500-8800 \AA \ range in a total of four clusters, one of them
([W99]2) common between ISAAC and UVES. For all 5 clusters we used I-band
HST imaging from Whitmore to derive half power radii from
King-model fitting. We find a range of cluster velocity dispersions from 9
to 20 km/s, and half power radii from 3.6 to 6 pc. In the cluster common to
UVES and ISAAC, both data sets are in excellent agreement and yield
velocity dispersions of 14.3 and 14.0 km/s, respectively. Masses are
derived from the Viral Theorem, which should be applicable since the
clusters obviously already have survived for 20-50 crossing times. The
corresponding dynamical masses range from 6--50~$\times$~10$^5$ M$_\odot$. 
Our data thus unambiguously show that the brightest Antennae clusters are 
indeed massive, with masses well above typical globular clusters 
($> 10^5$ M$_\odot$). Our results are in good agreement with the
hypothesis that the Antennae star clusters are globular clusters in
formation, given that the observed clusters are at the top of the Antennae
cluster luminosity function, that the HST imaging may somewhat overestimate
cluster sizes (and thus masses) and that stellar evolution will remove mass
from the clusters. Comparison with the N-body simulations of [32]
indicates that the Antennae clusters should survive
for several Gyrs but lose a significant fraction of their present mass.

In addition to the derived dynamical masses, we determined from our data the
ages (8-10 Myr: from CO/CaT and HI Br$\gamma $ equivalent widths) and the
V-band/K-band luminosities of the 5 clusters. Taken together, our
measurements constrain the distribution function of stellar masses in
the clusters, which should be close to their birth/Initial Mass Function
(IMF). The bottom inset in Figure 4 shows the results for four of our
clusters in the plane L$_{V}$/M vs log(t). For comparison we also plot the
positions of a cluster in the starburst galaxy M82 and two clusters in the
blue compact dwarfs NGC1705 and NGC1569 [30, 29].
Also plotted are L$_{V}/M$(t) ratios for power law IMFs with
different slopes $\alpha $, assuming a mass range from 0.1 to 100 M$_\odot$. 
A Salpeter IMF (solid line in Figure 4) has $\alpha $=2.35, and a
Salpeter IMF with a mass range of 1-100 M$_\odot$ has an L/M ratio that
is a factor of 2.6 greater. At face value the data of the different clusters
in Figure 4 require different IMF slopes (see also the analysis of the IMF
in the Galactic star cluster NGC 3603 [7]. Most
clusters are consistent with a Salpeter IMF, but (three of) the Antennae
clusters require a steeper IMF, and the M82 and NGC 3603 clusters require a
shallower IMF. To be consistent with a common (Salpeter-like) IMF, all
authors would have had to significantly underestimate the errors in the
analysis of their data. It is interesting to note that the clusters
requiring a steeper IMF are all found in the dusty `overlap region' between
NGC 4038 and NGC 4039. Figure 8 thus provides the tantalizing result that
IMFs vary in different environments. Obviously it is of great interest to
increase the statistics and confidence of this potentially far-reaching
result.

\section{First Results of an IR Tully-Fisher Study\protect\newline 
of Star Forming Disk Galaxies at z$\sim$1}

\begin{figure}[p]
\begin{center}
\includegraphics[width=10cm, bb = 68 25 545 720, clip=true]{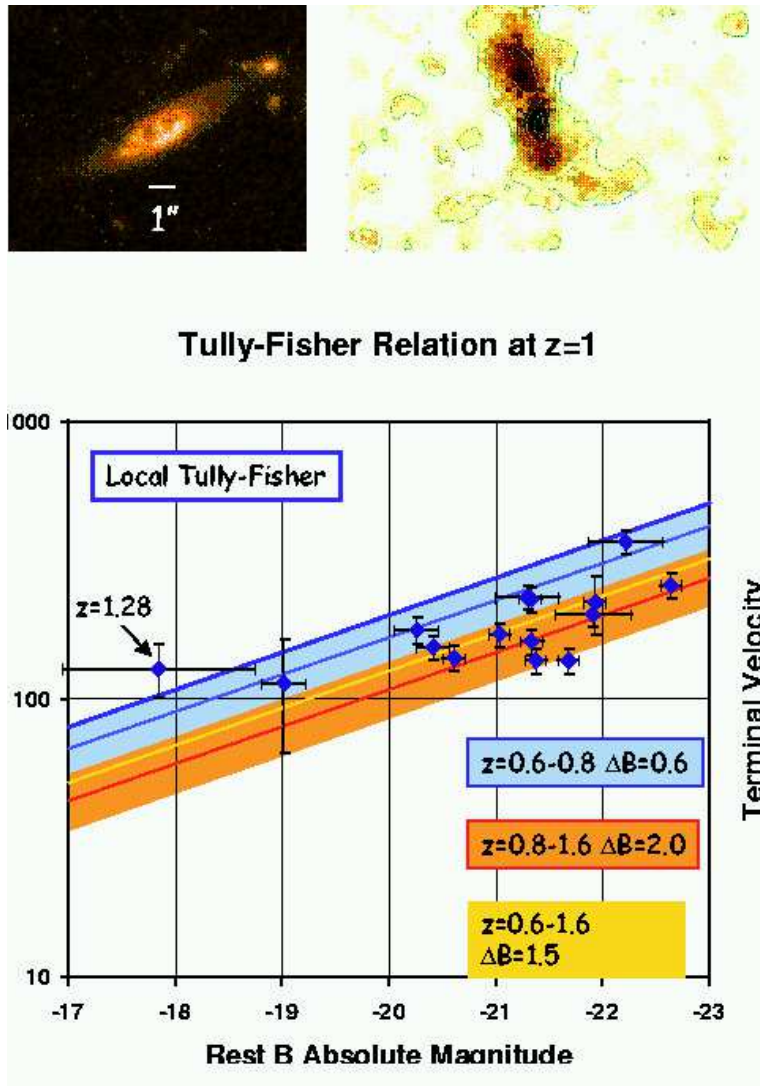}
\end{center}
\caption[]{Tully-Fisher diagram for 15 disk galaxies between z=0.6 and 1.6,
derived from ISAAC H$\alpha $ rotation curves. The
local Tully-Fisher curve is denoted by a thick blue line. The z\symbol{126}1
data are consistent with the same slope as that of the local curve, but
appear to require a brightening of 1.5 mag in the rest-B band}
\label{eps5}
\end{figure}

As part of the Ph.D. thesis of one of us (M.B.), we have recently begun a
program of near-IR spectroscopy of distant disk galaxies for a determination
of H$\alpha $ rotation curves at z$\sim$1. Our program extends to
higher redshift earlier [OII] studies at z$\sim$0.4--0.9 undertaken by
the Lick group [34, 35, 18]. We have observed with
ISAAC 20 inclined disk galaxies, selected from the CFRS, Hawaii medium deep
and Caltech faint galaxy surveys to lie in the redshift range 0.6--1.6. HST
images and photometry are available for nearly all of these sources. We
gave preference to those systems with significant [OII] emission in the
existing optical spectra. So far we have been able to extract H$\alpha $
rotation curves for 16 galaxies. The top insets of Figure 5 show a typical
(HST) image and a position-velocity diagram for one of our galaxies. In that
case, the turnover of the observed rotation curve to the flat part appears
to be well sampled. In other cases, the derivation of the true rotation
curve requires careful modeling with an input rotation curve and taking into
account the spatial and spectral convolution of our finite resolution data
(see [34,35] for a discussion). A first cut Tully-Fisher
curve is shown in the lower part of Figure 5 (for a $\Omega _{m}=0.3,\Omega
_{\Lambda }=0.7$,h=0.7 cosmology). Our data are consistent with the same
slope as the local Tully-Fisher relationship but indicate a 1.5 mag
brightening in the rest-wavelength B-band. For comparison [35]
did not find any significant evolution of the Tully-Fisher relation at 
z$\sim$0.5 ($\Delta B\leq 0.6$ mag), while [23] and [28]
find a brightening of $\sim$1.5-2 mag at z$\sim$0.2--0.3
for blue, compact emission line galaxies. Our sample contains mainly large
disk galaxies (disk scale lengths of 5$\pm$3 kpc), similar to [35], perhaps 
indicating a more
significant evolution at the higher redshifts we are sampling (see also
Weiner, this symposium). We are presently investigating in more detail the
possible influence of the selection criteria of our sample on the results.

\end{document}